\RequirePackage[2020-02-02]{latexrelease}
\documentclass[aps,prd]{revtex4}
\usepackage[colorlinks=true, pdfstartview=FitV, linkcolor=blue, citecolor=red, urlcolor=magenta]{hyperref}
\usepackage{graphicx}
\usepackage{latexsym}
\usepackage{amsmath}
\usepackage{amsfonts}
\usepackage{amssymb}
\usepackage{verbatim}

\newcommand{\be}{\begin{equation}}
\newcommand{\ee}{\end{equation}}
\newcommand{\bea}{\begin{eqnarray}}
\newcommand{\eea}{\end{eqnarray}}

\newcommand{\bb}{\bibitem}

\def\bb{\bibitem}

\def\bb{\bibitem}
\newcommand{\ben}{\begin{eqnarray}}
\newcommand{\een}{\end{eqnarray}}
\newcommand{\sech}{\rm sech}

\begin{document}


\title{The physics of intersecting thick to thin branes}

\author{$^{1}$Jos\'e L. Paulino}
\email{leonardopaulino\_pl@hotmail.com}
\author{ $^{1,2}$Francisco A. Brito}
\email{fabrito@df.ufcg.edu.br}

\affiliation{$^{1}$Departamento de F\' isica, Universidade Federal da Para\' iba,\\  Caixa Postal 5008, Jo\~ ao Pessoa, Para\' iba, Brazil.}
\affiliation{$^{2}$Departamento de F\'{\i}sica, Universidade Federal de Campina Grande,\\
Caixa Postal 10071, 58429-900, Campina Grande, Para\'{\i}ba, Brazil.}




\begin{abstract}
We model four-dimensional junctions made out of intersections of co-dimension one brane living in higher-dimensional spacetimes through domain walls.  We take a new look at the problem of localizing fermion states on brane junctions as a result of intersecting one thick brane to others sufficiently thin. All the branes intersect orthogonally to form a four-dimensional junction embedded in a higher-dimensional bulk. We discuss the effects of the Yukawa coupling and the proton decay on the restriction of the parameters that control the junction stability and the brane thickness which also define the bulk cosmological constant.
\end{abstract}
\pacs{XX.XX, YY.YY} \maketitle


\section{Introduction}

Intersecting branes have addressed in the literature with several distinct interests, such as dimensional reduction, as a mechanism of localizing gravity in arbitrary number of infinite dimensions \cite{ADDK2000,FR2006, CS2000},  dimensionality selection of the Universe \cite{BBL2006} and particle phenomenology in which several configurations of D-brane intersections have been considered to address standard model particles in string theory --- see \cite{Aldazabal:2000cn,Cvetic:2011vz} and references therein.

Inspired by the scenarios \cite{ADDK2000} and \cite{BBF2009} with thin and thick branes, respectively,  we consider the set-up of intersecting $N$ orthogonal $N+2$-spatial dimensional branes in a  $(3+N)+1$-dimensional bulk with a certain cosmological constant. The $(N+2)$-branes are co-dimension one defects. In the present study these branes can have arbitrary thickness that enjoy several relationships among them. They will serve as mechanism to address several phenomenological issues such as fermion hierarchy and entropy of the system. The limit where the entropy is maximized at the junction, corresponds to the `thin wall limit'  which characterizes an effective compactification mechanism.

The brane thickness allows to study mass hierarchies of leptons in the standard model without symmetries \cite{AS2000} since the Yukawa couplings are governed by the overlap of fermion wave functions localized at different positions along the brane thickness. Same should be said for standard model particles living in a junction that results from brane intersections. Here we allow that intersecting branes  can assume different thickness such that the junction is effectively four-dimensional but with a thickness extending along with an extra dimension.  We have seen that in order to respect mass hierarchies the brane thickness should also respect its own hierarchy. For instance, to respect the proton decay, at least one of the branes that intercepts in a junction should be at least 40 times as much larger than the other thick branes. Thus, there is a lower bound that saturates precisely in the limit of the proton decay. There are two fundamental parameters in the theory that are constrained by the proton decay and bulk cosmological constant. They are the parameter that control the junction stability and the brane thickness. In the present study we focus ourselves on orthogonal intersection of branes of co-dimension one, i.e., domain walls type defects. We consider two possibilities, namely, intersections of six 8-branes in ten-dimensional spacetime and three 5-branes in seven-dimensional spacetime.

The paper is organized as follows. In Sec.~\ref{sec01} we present our model and find the domain wall solutions due to the presence of $N$ scalar fields and show how they can intersect to form junctions. In Sec.~\ref{sec02} we show how the fermions with different masses can be localized on the junction made out of orthogonal intersection of six 8-branes in ten-dimensional spacetime. The consideration of brane and junction thickness on the maximal entropy of the system is also addressed. In Sec.~\ref{sec03} we consider the scenario of orthogonal intersection of three 5-branes in seven-dimensional spacetime to address the Yukawa coupling and the proton decay. Finally in Sec.~\ref{sec04} we make our final comments.

\section{A model with soft supersymmetry breaking terms}\label{sec01}

\subsection{The $N$ scalar fields superpotential}
\begin{figure}[h!]
\begin{center}
  \includegraphics[scale=0.7]{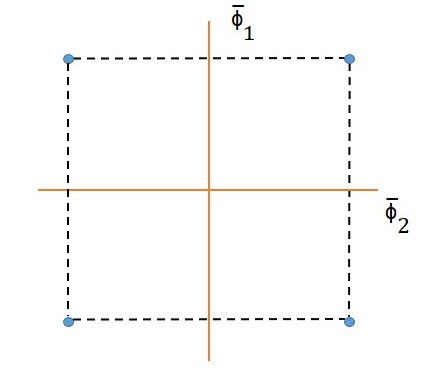}
  \caption{Two fields standard configuration.}
  \label{fig.1}
  \end{center}
\end{figure}

\begin{figure}[h!]
\begin{center}
\includegraphics[scale=0.2]{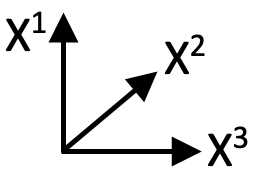}
  \includegraphics[scale=0.42]{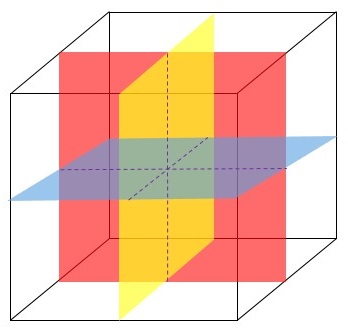}
  \includegraphics[scale=0.2]{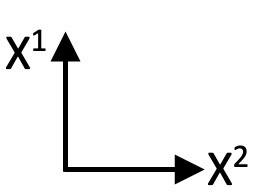}
   \includegraphics[scale=0.5]{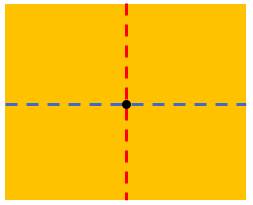}
  \caption{Three fields standard configuration.}
  \label{fig.2}
  \end{center}
\end{figure}

Let us start by introducing our supersymmetric Lagrangian \cite{BBL2006}
\begin{equation}
{\cal L}=\frac{1}{2}{\partial_m\phi^i}{\partial^m\phi^i}+\bar{\psi^i}\Gamma^m\partial_m\psi^i+W_{\phi^i\phi^j}\bar{\psi^i}\psi^j-V(\phi^i)-\frac{1}{2}{\varepsilon F(\phi^i)},
\label{eq2.0}
\end{equation}
where $m = 0, 1, 2, ..., D - 1$ and $i, j = 1, 2, ..., N$ transverse coordinates. The scalar potential $V$ is given in terms of the superpotential $W$ \cite{superpotencial1} \cite{superpotencial2} \cite{superpotencial3} \cite{superpotencial4} \cite{superpotencial5} \cite{superpotencial6}, so that
\begin{equation}
V(\phi^1,\phi^2,...\phi^N)=\frac{1}{2}\left({\frac{\partial W}{\partial\phi^1}}\right)^2+\frac{1}{2}\left({\frac{\partial W}{\partial\phi^2}}\right)^2+...+\frac{1}{2}\left({\frac{\partial W}{\partial\phi^N}}\right)^2.
\label{eq2.0a}
\end{equation}
From \eqref{eq2.0} and \eqref{eq2.0a}, we can write the effective potential as
\begin{equation}
V_{\varepsilon}=\frac{1}{2}\sum^N_{i=1}(\partial_{\phi_i}W)^2+\frac{1}{2}\varepsilon F(\phi^1,\phi^2,...,\phi^N),
\label{eq2.1}
\end{equation}
where $\varepsilon$ is a small parameter that plays an importante role on the stability of the junctions, $W$ is the superpotential and $F(\phi^1,\phi^2,...,\phi^N)$ is a function representing the combination of $N$ fields in the form $F(\phi^i)=\sum^{N}_{j>i}F(\phi^i,\phi^j)$, which in our case is given by
\begin{equation}
F(\phi^i,\phi^j)=\frac{1}{2}(\phi^{i^4}+\phi^{j^4})-3\phi^{i^2}\phi^{j^2}+\frac{9}{2}.
\label{eq2.1a}
\end{equation}
Let us consider the superpotential given in the form \cite{BBF2009}
\begin{equation}
W(\phi_i)=\sum^N_{i=1}\lambda_i\left(\frac{\phi^{3}_i}{3}-a^{2}\phi_i\right).
\label{eq2.2}
\end{equation}
This superpotential guarantees that the minima of the potential (\ref{eq2.1}) are always equal and equidistant, so that
\begin{equation}
\bar{\phi}_1=\bar{\phi}_2=...=\bar{\phi}_N=\pm\sqrt{\frac{3}{2-3(N-1)\varepsilon}},
\label{eq2.3}
\end{equation}
where we have written $a=u$ and $\lambda_i=-1/u$ in terms of a single parameter $u=\sqrt{3/2}$. 
For two scalar fields, for example, the minima are the following
\begin{equation}
\bar{\phi}_1=\bar{\phi}_2=\pm\sqrt{\frac{3}{2-3\varepsilon}}.
\label{eq2.4}
\end{equation}
For this type of solution, with equidistant minima, we have a configuration that is depicted in Fig.\ref{fig.1}, where two `1-brane' domain walls (orange solid lines) orthogonally intersect to form a `0-brane' junction. 
The points setting the vertices of the square are the minima of the potential. 
If we extend the previous example for three fields, we will find again equidistant minima, but now the configuration will be three-dimensional, 
where three `2-brane' domain walls orthogonally intersect to form a `0-brane' junction living in three spatial dimensions --- See two-dimensional slice in Fig.~\ref{fig.2} (right). Here, the minima of the potential are the points setting the vertices of a regular cube ---Fig.~\ref{fig.2} (left). 

The domain wall junction seen in Fig.\ref{fig.1} 
resulted from a configuration where the minima of the potential are equidistant, which means that the intersecting domain walls have the same (thin) width. The configuration of three fields will be presented later as the ideal setup to consider three 5-branes intersecting in right angles at 6+1 dimensions. The `0-brane' junction at such a seven-dimensional bulk is translated to a `3-brane' junction.

\subsection{The new scenario: modified superpotential}

Once we have presented our main ideas to address the scenario of intersecting branes, we will now look for a different type of junction. We will present a different configuration where the minima of the potential is no longer equidistant, which will result in domain walls that may not have the same thickness. Such junction will come from a modification of the superpotential \eqref{eq2.2}, which will result in a form that is much richer than that we have previously seen, i.e., we shall be now intersecting thick to thin branes. As we shall see, this completely modifies the physics of a junction made out of intersection of branes with same (thin) thickness. 

So, let us consider the same superpotential general form  (\ref{eq2.2}), i.e., 
\begin{equation}
W(\phi_i)=\sum^N_{i=1}\lambda_i\left(\frac{\phi^{3}_i}{3}-a_i^{2}\phi_i\right),
\label{eq2.5}
\end{equation}
but now we have $\lambda_1=\lambda_2=...=\lambda_N$ and $a_1=ka_2=ka_3...=ka_N$. Here, $k$ is a real number defined in the interval $(0,1)$, so that $\Delta_1\simeq1/\lambda_1a_1=1/k\lambda_2a_2$, where $\Delta_1$ is the width of the brane which is between the minima $\pm\bar{\phi}_1$. In other words, the lower is $k$, the larger is the width of the brane.

The calculation of the minima of the potential (\ref{eq2.1}), for $\phi_1$, $\phi_2$ and $\phi_3$, results in \begin{eqnarray}
\phi_{{1}}=\frac12\,{\frac {\sqrt { \left( 45\,{\varepsilon}^{2}-3\,\varepsilon-
4 \right)  \left( -18\,\varepsilon+3\,{k}^{2}\varepsilon-4\,{k}^{2} \right)
}\sqrt {6}}{45\,{\varepsilon}^{2}-3\,\varepsilon-4}},
\label{eq2.6}
\end{eqnarray}
\begin{eqnarray}
\phi_{{2}}=\frac12\,\sqrt {-{\frac {54\,{k}^{2}\varepsilon+36\,\varepsilon+24}{
45\,{\varepsilon}^{2}-3\,\varepsilon-4}}},
\label{eq2.7}
\end{eqnarray}
\begin{eqnarray}
\phi_{{3}}=\frac12\,\sqrt {-{\frac {54\,{k}^{2}\varepsilon+36\,\varepsilon+24}{
45\,{\varepsilon}^{2}-3\,\varepsilon-4}}},
\label{eq2.8}
\end{eqnarray}
where we have considered $\lambda_i=-1/u$, $a_1=ku$ and $a_2=a_3=a_4...=u$.

\begin{figure}[h!]
\begin{center}
  \includegraphics[scale=0.7]{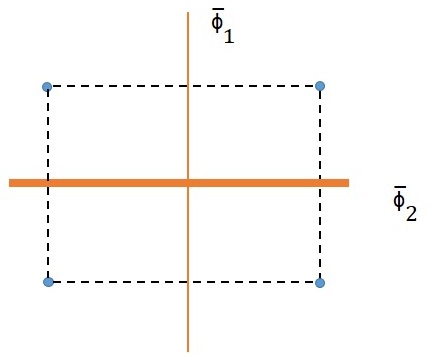}
  \caption{Two fields configuration as a result of the superpotential modification.}
  \label{fig.1.3}
  \end{center}
\end{figure}

These solutions show us that now the minima of the potential are no more equidistant points. The configuration of the junction in this new scenario for two fields is shown in Fig.~\ref{fig.1.3}.
\begin{figure}[h!]
\begin{center}
\includegraphics[scale=0.2]{axis-xyz-2}
  \includegraphics[scale=0.22]{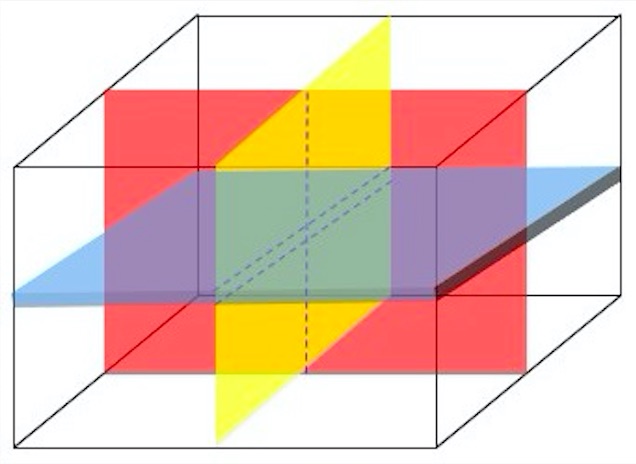}
  \includegraphics[scale=0.2]{axis-xy}
   \includegraphics[scale=0.5]{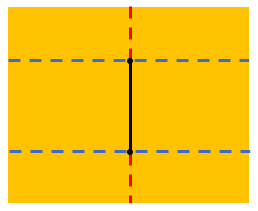}
  \caption{Three fields configuration as a result of the superpotential modification.}
  \label{fig.1.4}
  \end{center}
\end{figure}

Now, we no longer have two simple 1-brane intersecting to form a 0-brane, but instead, the domain walls has an extra dimension, which results from the minima $\pm\bar{\phi}_{{1}}$ getting closer. 
Due to the modified superpotential, the horizontal line in Fig.\ref{fig.1} becomes a `stripe' with certain width. 
It means that the intersection of the two domain walls no longer results in a simple 0-brane, but creates a brane with an extra dimension.
For three fields, the configuration develops a similar phenomenon. 
In the previous scenario, we had a configuration with three 2-branes, embedded in a three-dimensional cube, intersecting to form a 0-brane. Now, we no longer have a regular cube --- See Fig.~\ref{fig.1.4} (left). Again, with one of the domain walls having an extra dimension, the junction is no longer a simple 0-brane, but rather we have a higher dimensional `dimer-like' structure, living in three spatial dimensions --- See Fig.~\ref{fig.1.4} (right). 

\begin{figure}[h!]
\begin{center}
  \includegraphics[scale=0.3]{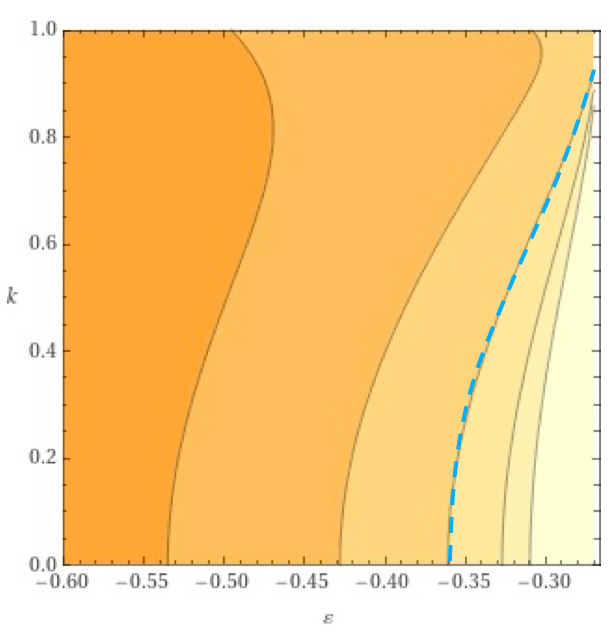}
  \caption{The contour plot of $\Lambda(k,\varepsilon)$ given in Eq.~(\ref{eq2.9}). The contour along the dashed-blue curve passing through $\varepsilon\approx-0.36$ is an importante dividing line, i.e., along with this line $\Lambda=0$, whereas in the regions inside, $\Lambda<0$ and outside $\Lambda>0$. $\Lambda$ diverges at $\varepsilon=-4/15\approx -0.2667$.}
  \label{fig.1.3.1}
  \end{center}
\end{figure}

It is also important to show that, in our present scenario, the cosmological constant $\Lambda\equiv V_{\varepsilon}(\bar{\phi}_1,\bar{\phi}_2,...,\bar{\phi}_N)$ can be negative, as it is shown in \cite{BBF2009}, since $\varepsilon < 0$. For the new superpotential, the vacua of the potential are given by \eqref{eq2.6}, \eqref{eq2.7} and \eqref{eq2.8}, so that
\begin{eqnarray}
\Lambda=\frac94\,{\frac {\varepsilon\, \left( 135\,{\varepsilon}^{2}+21\,\varepsilon+15\,{k
}^{4}\varepsilon+12\,{k}^{2}-2\,{k}^{4}-10 \right) }{45\,{\varepsilon}^{2}-3
\,\varepsilon-4}},
\label{eq2.9}
\end{eqnarray}
for three fields. It has been shown that $\Lambda$ is always negative \cite{ADDK2000} \cite{CS2000} \cite{CHT2000} \cite{O2000} \cite{ST2001} \cite{FR2006}. However, because of the dependence on the two parameters, $\varepsilon$ and $k$, 
now $\Lambda$ develops negative, zero or positive values at certain regions. See the contour plot in  Fig.~\ref{fig.1.3.1}. This is will lead to interesting phenomenology that we shall address later. 

\begin{figure}[h!]
\begin{center}
 \includegraphics[scale=0.2]{axis-xy}
  \includegraphics[scale=0.3]{fig1-190822}
  \includegraphics[scale=0.3]{fig2-190822}
  \includegraphics[scale=0.3]{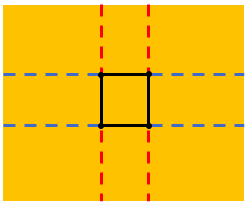}
  \includegraphics[scale=0.3]{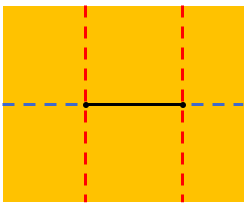}
\caption{The planar representations of the intersection of three domain walls depicted in Fig.~\ref{fig.2} and Fig.~\ref{fig.1.4}. They are named slices (a), (b), (c) and (d) from the left to the right, that are seen by an observer sitting along the $x^3$-axis. 
As a result of the superpotential modification we may find: (a) only thin branes intersects, (b) one brane becomes thicker along $x^1$, (c) two branes become thicker along $x^1$ and $x^2$, finally (d) the brane thickness along the coordinate $x^1$ shrinks, whereas one brane keeps thicker along $x^2$.}
\label{fig.1.4.1}
 \end{center}
\end{figure}

Before ending this section, some important comments are in order. The above scenario, where we consider the thickness involving just one of the intersecting branes, can be extended to any number of thicker branes. We shall investigate this possibility, as we shall address the issue of interplaying the thickness of two intersecting branes. For instance, we shall show that there is and interesting phenomenon by 
interchanging the role of the thickness $\Delta_1$ with $\Delta_2$. This is depicted in Fig.~\ref{fig.1.4.1} as a transition that leaves the configurations (a) to (d). In particular, as we transit from configuration (b) to (d) we can perform the duality transformation
 \begin{eqnarray}
 \Delta_2\to \Delta_1,\qquad  \Delta_1\to \Delta_2,  
\qquad k\to 1/k,
 \label{dual}
 \end{eqnarray}
without changing the energy of the junction. However, this duality symmetry certainly is not respected by the cosmological constant (\ref{eq2.9}), except along the contour where $\Lambda=0$ --- See Fig.~\ref{fig.1.3.1}. Furthermore, since we are interchanging different geometric aspects of the theory, this transformation is analogous to the T-duality \cite{Alvarez:1994dn}. 
We shall return to further discussions on this issue shortly.

\section{Fermion states and the energy on the junction}\label{sec02}

The equations of motion for bosonic and fermionic fields resulting from \eqref{eq2.0} are
\begin{equation}
\Box\phi^i+\frac{\partial V}{\partial\phi^i}-W_{\phi^i\phi^i\phi^i}\bar{\psi^i}\psi^i+\frac{\varepsilon}{2}\frac{\partial F}{\partial\phi^i}=0,
\label{eq3.0}
\end{equation}
\begin{equation}
\Gamma^m\partial_m\psi^i+W_{\phi^i\phi^i}\psi^i=0.
\label{eq3.0a}
\end{equation}
In the presence of a domain walls \cite{naosei1} \cite{naosei2} \cite{naosei3} \cite{naosei4} \cite{naosei5}, the equation of motion \eqref{eq3.0a} can be written in terms of $ \psi^{i}_\pm$. The fermionic solutions, in this case, are given in the form
\begin{equation}
\psi^{j}=e^{ip^{(j)}_{\mu}x^\mu}\chi^{j}(x^j),
\label{eq3.0b}
\end{equation}
where $p^{(j)}_{\mu}$ is the momentum along the $j$-th domain walls whose indices $\mu = 1, 2, 3, ..., D - 2$ label the coordinates on its world-volume. By considering our system in a rest frame and using \eqref{eq3.0b} into \eqref{eq3.0a}, we find
\begin{equation}
iE^j\Gamma^{0}\chi^{j}+\Gamma^{j}\partial_j\chi^{j}+W_{\phi^j\phi^j}\chi^{j}=0.
\label{eq3.0c}
\end{equation}
If we use the properties of the gamma matrices, $\Gamma^j\chi_\pm=\pm\chi_\pm$ and $i\Gamma^0\chi_\pm=\chi_\mp$, we can show that \eqref{eq3.0c} leads us to the Schrödinger-like equations
\begin{equation}
[-\partial^{2}_j+U^{j}_\mp(x^j)]\chi^{j}_\pm=E^{2}_j\chi^{j}_\pm,
\label{eq3.0d}
\end{equation}
with
\begin{equation}
U^{j}_\mp(x^j)=W^{2}_{\phi^j\phi^j}(x^j)\mp W{'}_{\phi^j\phi^j}(x^j).
\label{eq3.0e}
\end{equation}

These equations govern the dynamics of fermion bound states associated with independent domain walls. Therefore, to describe domain walls junctions, we need to consider the Schrödinger-like equation in the form
\begin{equation}
[-\nabla^2+U_{junc}]\psi_{\pm(n_1...n_N)}=E^{2}_{(n_1...n_N)junc}\psi_{\pm(n_1...n_N)},
\label{eq3.0f}
\end{equation}
where
\begin{equation}
U_{junc}=U^{1}_\mp(x^1)+U^{2}_\mp(x^2)+...+U^{N}_\mp(x^N),
\label{eq3.0g}
\end{equation}
\begin{equation}
E^{2}_{(n_1...n_N)junc}=E^{2}_{(n_1)1}+E^{2}_{(n_2)2}+...+E^{2}_{(n_N)N},
\label{eq3.0h}
\end{equation}
\begin{equation}
\psi_{\pm(n_1...n_N)}=\chi^1_{\pm(n_1)}(x^1)\times\chi^2_{\pm(n_2)}(x^2)\times...\times\chi^N_{\pm(n_N)}(x^N),
\label{eq3.0i}
\end{equation}
with the components $\chi^i_{+(n_i)}(x^i)$ and $\chi^i_{-(n_i)}(x^i)$ being normalizable functions and $n_i=0,1$ labels the two bound states that can be found both on individual domain walls and on the junction. Only one of these components is normalizable for the zero mode, i.e., for the state where $E=0$, which describes a chiral fermion. The tower of continuum states are non-normalizable modes and cannot be localized to the defects, but they can be accessed at sufficiently high energy.

Let us start by considering the example with $N = 2$ scalar fields forming walls of independent domains, which come together to form a junction in a spacetime with $D = 3 + 1$ dimensions. We shall now make use of the modified superpotential, so that we have 
\begin{equation}
W(\phi_1,\phi_2)=\lambda_1\left(\frac{\phi^{3}_1}{3}-a_1^{2}\phi_1\right)+\lambda_2\left(\frac{\phi^{3}_2}{3}-a_2^{2}\phi_2\right),
\label{eq3.0j}
\end{equation}
where, as aforementioned, $a_1^2=k^2a_2^2$.

At sufficiently small symmetry breaking parameter $\varepsilon$, the Eqs.~\eqref{eq3.0}-\eqref{eq3.0a}, are satisfied by static BPS domain walls that solve the following first order equations
\begin{equation}
\frac{d\phi^i}{dx^i}=\frac{\partial W}{\partial\phi^i}\,,\qquad i=1,2,...,N.
\label{eq3.0k}
\end{equation}
So, for two scalar fields, the equation \eqref{eq3.0k} becomes
\begin{equation}
\frac{d\phi^1}{dx^1}=\frac{\partial W}{\partial\phi^1},\qquad \frac{d\phi^2}{dx^2}=\frac{\partial W}{\partial\phi^2},
\label{eq3.3}
\end{equation}
whose solutions are
\begin{equation}
\phi^1(x^1)=-a_1\tanh(\lambda_1a_1x^1),\,\,\,\,\,\,\,\,\,\,\phi^2(x^2)=-a_2\tanh(\lambda_2a_2x^2).
\label{eq3.4}
\end{equation}
By replacing \eqref{eq3.0j} into \eqref{eq3.0e} and considering the solutions given by \eqref{eq3.4}, we obtain the following potentials
\begin{equation}
U^1_+=4\lambda^2_1a_1^2-6\lambda^2_1a_1^2\,{\sech}^2(\lambda_1a_1x^1),
\label{eq3.5}
\end{equation}
\begin{equation}
U^2_+=4\lambda^2_2a_2^2-6\lambda^2_2a_2^2\,{\sech}^2(\lambda_2a_2x^2).
\label{eq3.6}
\end{equation} 
They are modified Pöschl-Teller potentials \cite{pochete}, which are given by the general form $U(x^i)=A-B\, {\sech}^2(x^i)$ for $i=1,2$, and $A, B$ are real constants. The normalized bound states have the following energies
\begin{equation}
E_n=A-\left[\sqrt{B+\frac{1}{4}}-\left(n+\frac{1}{2}\right)\right]^2,
\label{eq3.2}
\end{equation}
where
\begin{equation}
n=0,1,...<\sqrt{B+\frac{1}{4}}-\frac{1}{2}.
\label{eq3.2a}
\end{equation}
Replacing the equations \eqref{eq3.5} and \eqref{eq3.6} into \eqref{eq3.0d}, we obtain
\begin{equation}
\left\{-\frac{d^2}{d{y^{j}}^2}+\Big[4-6{\,\sech}^2(y^{j})\Big]\right\}\chi^{j}_-=\frac{E^{2}_{j}\chi^{j}_-}{\lambda^2_{j}a_{j}^2},
\label{eq3.2b}
\end{equation}
where we have assumed the scaling $\lambda_{j}a_{j}x^{j} \rightarrow y^{j}$. 

Now, comparing the above equation with the general form of the Pöschl-Teller potential and using \eqref{eq3.2} and \eqref{eq3.2a}, we determine that the discrete spectrum is composed by two bound states, the zero mode and one excited bound state, so that
\begin{equation}
{E^{2}_{(0)}}_{1,2}=0,
\label{eq3.7}
\end{equation}
\begin{equation}
\chi^{1,2}_{(0)}=C_0{\sech}^2(\lambda_{1,2}a_{1,2}x^{1,2}),
\label{eq3.8}
\end{equation}
\begin{equation}
{E^{2}_{(1)}}_{1,2}=3\lambda_{1,2}^2a_{1,2}^2,
\label{eq3.9}
\end{equation}
\begin{equation}
\chi^{1,2}_{(1)}=C_1\tanh(\lambda_{1,2}a_{1,2}x^{1,2}){\sech}(\lambda_{1,2}a_{1,2}x^{1,2}).
\label{eq3.10}
\end{equation}

This is the spectrum referring to fermions bound to the domain walls. The spectrum associated with a domain wall junction is obtained considering the equation \eqref{eq3.0f}. Notice that there are four possible combinations by using the zero mode and the excited bound state, given by
\begin{equation}
{E^{2}_{(00)}}_{junc}=0,
\label{eq3.11}                        
\end{equation}
\begin{equation}
\psi_{(00)}=C_1{\sech}^2(\lambda_{1}a_1x^{1})\times{\sech}^2(\lambda_{2}a_2x^{2}),
\label{eq3.12}
\end{equation}
\begin{equation}
{E^{2}_{(01)}}_{junc}=3\lambda^{2}_{2}a_2^2\equiv 3,
\label{eq3.13}
\end{equation}
\begin{equation}
\psi_{(01)}=C_2{\sech}^2(\lambda_{1}a_1x^{1})\times\tanh(\lambda_{2}a_2x^{2})\times{\sech}(\lambda_{2}a_2x^{2}),
\label{eq3.14}
\end{equation}
\begin{equation}
{E^{2}_{(10)}}_{junc}=3\lambda^{2}_{1}a_1^2\equiv3\, k^2,
\label{eq3.15}
\end{equation}
\begin{equation}
\psi_{(10)}=C_3\tanh(\lambda_{1}a_1x^{1})\times{\sech}(\lambda_{1}a_1x^{1})\times{\sech}^2(\lambda_{2}a_2x^{2}),
\label{eq3.16}
\end{equation}
\begin{equation}
{E^{2}_{(11)}}_{junc}=3\lambda^{2}_{1}a_1^2+3\lambda^{2}_{2}a_2^2\equiv3(1+k^2),
\label{eq3.17}
\end{equation}
\begin{equation}
\psi_{(11)}=C_4\tanh(\lambda_{1}a_1x^{1})\times{\sech}(\lambda_{1}a_1x^{1})\times\tanh(\lambda_{2}a_2x^{2})\times{\sech}(\lambda_{2}a_2x^{2}).
\label{eq3.18}
\end{equation}
where, for the sake of simplicity, in Eqs.~\eqref{eq3.13}, \eqref{eq3.15} and \eqref{eq3.17}, we considered $\lambda_i=-1/u, a_1=k a_2=...=k a_N, a_2=a_3=...=a_N=u$, adopted in Sec.~\ref{sec01}. Notice that, for $k=1$, these results for both the fermion spectrum and the junction spectrum are the same as those obtained in  \cite{BBF2009}. 
The fermionic spectrum above is the same spectrum for the bosonic scalar sector. Furthermore, we can easily check the duality transformation (\ref{dual}) by simply verifying the equation (\ref{eq3.17}).
\section{Counting the number of bound states}

In the present study the Schr\"odinger-like potentials admit two bound states separated from the continuum by an energy gap. For sufficiently low energy, our system can essentially be treated as a two level system. Two level systems are well-known to be related to the existence of a point of maximal entropy. For $M_1$ particles at the lowest bound state and $M_2$ particles at the highest bound state, the number of accessible microstates by a two levels system follows the combinatorial 
\begin{equation}
\Omega=\frac{M!}{M_1!M_2!},
\label{distribution}
\end{equation}
where $M=M_1+M_2$. We have that the summation of the number of bound states is given by $\sum_{M_1=0}^M\Omega(M_1)=2^M$. For a junction made out of $N$ intersecting $(N+2)$-branes of co-dimension one in $D=d+1$ dimensions,  we have $2^M$ states as $M={\cal N}N$, where ${\cal N}$ is the number of  brane junctions in the bulk, normally admitted to be large. 
The entropy can be computed in terms of the number of accessible microstates according to the Boltzmann formula $S = \ln \Omega$.

By assuming the  typical `intermolecular' distance in the bulk is of the order of the branes thickness,  the number of accessible microstates can be given as follows:
\begin{equation}
M=\frac{V_d}{\Delta_1\Delta_2\cdot \cdot \cdot\Delta_{d-1}\Delta_d},
\label{states-number}
\end{equation}
where $V_d$ can be decomposed into $V_d=L^{d-1}r$, with  $r=\sqrt{x_1^2+x_2^2+...+x_{d}^2}$. 
We find that $r/\Delta_1$ is given by
\begin{equation}
\sqrt{\frac{x_1^2}{\Delta_1^2}+\frac{x_2^2}{\Delta_1^2}+...+\frac{x_{d}^2}{\Delta_1^2}}\leq \frac{x_1}{\Delta_1}+\frac{x_2}{\Delta_1}+...+\frac{x_{d}}{\Delta_1},
\label{distances-norm}
\end{equation}
where the triangle inequality is displayed here for later use. Thus, for $r$ projected to a plane, say $x_1=\Delta_1,\, x_2=x_3=...=x_d= \Delta_2$, then from Eq.~(\ref{states-number}) we find 
\begin{equation}
M=\frac{L^{d-1}}{\Delta_2\cdot \cdot \cdot\Delta_{d-1}\Delta_d}\sqrt{1+(d-1)\frac{\Delta_2^2}{\Delta_1^2}},
\label{states-number-2}
\end{equation}
which develops two asymptotic limits. Firstly, as $\Delta_2\ll\Delta_1=l_s$, for $\Delta_2=\Delta_3=...=\Delta_{d}=l_P$, the number of accessible microstates reads
\begin{equation}
M\sim S\propto\frac{L^{d-1}}{l_P^{d-1}}, 
\label{states-number-3}
\end{equation}
which coincides with the maximal entropy  --- see below --- that grows with the {\sl area} (the boundary) of the bulk rather than with the volume. On the other hand,  as $\Delta_2\gg\Delta_1=l_P$, for $\Delta_2=\Delta_3=...=\Delta_{d}=l_s$, we obtain
\begin{equation}
M=\frac{L^{d-1}}{l_s^{d-1}}\sqrt{d-1}\,\frac{\Delta_2}{\Delta_1}\sim S\propto\sqrt{D-2}\,M_s l_s,
\label{states-number-4}
\end{equation}
where in the last step we assumed the evolution of the entropy of the system to string entropy as $L\to l_s$, analogous to `stretched horizon' in the Schwarzschild black hole with temperature $T_H\sim 1/l_s$.

In the previous analysis we consider only the leading term of the entropy. Now, by expanding the entropy $S=\ln{\Omega}$ around the maximum $M_2=M/2$ we find the maximal entropy
\begin{equation}
S_{max}=\ln\left[\frac{\Gamma{(M+1)}}{\Gamma{(\frac{M+2}{2})^2}}\right], 
\label{entropy}
\end{equation}
from which we find the leading contribution with the next-to-leading correction. Thus, in the limit as $M\to\infty$ we can find the following 
\begin{eqnarray}
S_{max}&=&\ln{2}\; M-\frac{1}{2}\ln{M}+{\cal O}\left(\frac{1}{M}\right),
\label{s-entropy-0}
\end{eqnarray}
which $M$ is large in the both cases  Eq.~\eqref{states-number-3} and Eq.~\eqref{states-number-4}. Let us focus on the latter case, the string entropy, to obtain
\begin{eqnarray}
S_{max}\propto\sqrt{D-2}\,M_s l_s-\frac{1}{2\ln{2}}\ln{\,M_s l_s}.
\label{s-entropy-D}
\end{eqnarray}
Notice this is in accord with the entropy of a free string with subleading corrections, which is described by a 1+1 dimensional quantum field theory with $D-2$ transverse coordinates playing the role of free scalar fields living in the target space. 

Alternatively, from Eq.~\eqref{distances-norm} due to triangle inequality one can establish a new entropy upper bound. This is because $r/\Delta_1\leq n$, thus from Eq.~\eqref{states-number-4} by considering the norm $n$, we can reach the following result for the string entropy as $n\to \infty$
\begin{eqnarray}
\widetilde{S}_{max}&=&\ln{2}\; n-\frac{1}{2}\ln{n}+{\cal O}\left({\frac{1}{n}}\right)\nonumber\\
&=&\ln{2}\; \Big(k_2+k_3+...+k_{d}\Big)-\frac{1}{2}\ln{\Big(k_2+k_3+...+k_{d}\Big)},
\label{s-entropy}
\end{eqnarray}
where $n=1+k_2+k_3+..+k_{d}$ and $k_i=\Delta_{i}/\Delta_{1}$ are the ratio between the thickness of two intersecting branes --- one might want to understand $1/k_i$ as the analogous of the `winding number' --- see below.  In the example discussed in Sec.~\ref{sec01} with just one thicker brane, we had $k_2\equiv k$ and $k_i=1$, for $i>2$. This entropy may find some potential application in recents studies by considering lattice models, as for instance, the X-cube model \cite{Rudelius:2020kta}.


In the limit as $\Delta_i=\Delta_2$, $k_2=k_3=...=k_d=k$, then the Eq.~(\ref{s-entropy}) can be written as
\begin{eqnarray}
\widetilde{S}_{max}=\ln{2}\; \Big[(d-1)k\Big]-\frac{1}{2}\ln{\Big[(d-1)k\Big]}.
\label{s-entropy-2}
\end{eqnarray}
Since $k\simeq\Delta_2/\Delta_1$ was defined in the interval $(0,1)$, but Eq.~(\ref{s-entropy-2}) is consistent only for large $k$, we should `T-dualize' it according to the duality symmetry $k\to 1/k$, $\Delta_2\to\Delta_1$, $\Delta_1\to\Delta_2$ as given in Eq.~(\ref{dual}). This will provide us with the $k$ in the new interval $(1,\infty)$. This process allows a transformation such that  $\Delta_{2}\ll \Delta_{1}\to\Delta_{2}\gg \Delta_{1}$, which can be geometrically understood from Figs.~(\ref{fig.1.4.1}b) and (\ref{fig.1.4.1}d). Now mapping the winding mode as periodicity along the lattice in the coordinate $x^2$, Fig.~(\ref{fig.1.4.1}d), then $\Delta_2=2\pi \ell R$, $\Delta_1=2\pi l_s$ and we find
\begin{eqnarray}
\widetilde{S}_{max}&=&\ln{2}\; \left[(d-1)\frac{\ell R}{l_s}\right]-\frac{1}{2}\ln{\left[(d-1)\frac{\ell R}{l_s}\right]}, \nonumber\\
&\propto& (D-2)M_s l_s - \frac{1}{2\ln{2}}\ln{M_s l_s}.
\label{s-entropy-3}
\end{eqnarray}
In the last step we have used the minimum mass of a stretched string given by
\begin{equation}
M_s=\frac{\ell R}{\alpha'},
\label{Mstring}
\end{equation}
where ${\ell }$ is the winding number, $\alpha'=l_s^2$ and $R$ is the compactification radius along a transverse coordinate, say, $x^{2}(\tau,\sigma+2\pi)=x^{2}(\tau,\sigma)+2\pi\ell R$, where $(\tau,\sigma)$ are string world-sheet coordinates. Finally, from Eqs.~\eqref{s-entropy-D}-\eqref{s-entropy-3} we establish the inequality 
\begin{equation}
S_{max}\leq\widetilde{S}_{max}.
\label{entropy-bound}
\end{equation}
Concerning these computations, some comments are in order. The results presented here is 
indeed due to a connection between the brane thickness $\Delta_i$ and Unruh  temperature as $T_H\sim 1/\Delta_i\sim \lambda_i\, a_i$ according to a Rindler observer along the orbits connecting the vacua \cite{Brito:2012gp,Bazeia:2016pra} of a system described by the superpotential defined in Eq.~\eqref{eq2.5}, which are a collection of decoupled superpotentials. The relationship between coupled and decoupled superpotentials can be seen by properly rotating the fields and constraining the independent coupling constants as has been shown long ago \cite{Brito:1997ra,Bazeia:1998zv}. Thus, from Eq.~\eqref{states-number}, for $T\equiv T_H\sim1/\Delta_i$, with $\Delta_i=\Delta$ we have 
\begin{equation}
M\sim S\propto T^d\, V_d,
\label{states-number-T}
\end{equation}
which is the entropy density of a $d+1$ dimensional free scalar field theory. Let us now admit the the entropy is stored in a layer $\delta \Delta$ such that we can write
\begin{eqnarray}
\delta S&\propto& T^d\, L^{d-1}\delta\Delta,\nonumber\\
&\propto& L^{d-1}\frac{\delta\Delta}{\Delta^d}.
\label{states-number-dT}
\end{eqnarray}
By integrating in $\Delta$ we find the entropy
\begin{eqnarray}
S&\propto& L^{d-1}\int_\epsilon^\infty\frac{\delta\Delta}{\Delta^d}\propto\frac{L^{d-1}}{\epsilon^{d-1}}\nonumber\\
&\propto&\frac{L^{d-1}}{l_P^{d-1}},
\label{states-number-dT2}
\end{eqnarray} 
which is in agreement with Eq.~\eqref{states-number-3}. The UV cut-off $\epsilon$ was assumed to be of the order of the Planck length. Thus, the entropy is mainly localized around the horizon where the temperature becomes higher, i.e., $T\sim 1/l_P$. One of the interesting phenomena that we can see from Eq.~\eqref{states-number-dT2} is the fact that in the integral $\Delta$ spans from infinity to $l_P$, which characterizes a compactification, as expected from the {\sl holographic} perspective \cite{tHooft:1993dmi,Susskind:1994vu}.

\subsection{An example with six extra dimensions}

Now by considering a system as played by the intersection of six eight-dimensional domain walls representing `8-branes' living in ten dimensions, the scalar and fermion fields on the junction are obtained through the following spectral decomposition
\begin{align}
\phi-\phi_s & =\eta(y^{\mu};x_1,..,x_6) \nonumber \\ 
            & =\sum_{n_1...n_6}\xi^{junc}_{n_1...n_6}(y^{\mu})\psi^{n_1...n_6},
\label{eq3.18c}
\end{align}
\begin{equation}
\Psi(y^{\mu};x_1,..,x_6)=\sum_{n_1...n_6}\tau^{junc}_{n_1...n_6}(y^{\mu})\psi^{n_1...n_6},
\label{eq3.18d}
\end{equation}
where $n_i=0,1$ and $\psi^{n_1...n_6}=\chi^{n_1}(x^1)\times\chi^{n_2}(x^2)\times... \times\chi^{n_6}(x^6)$, and $\chi(x_i)$ are functions that satisfy the equation \eqref{eq3.0f}, valid for both fermions and bosons.
Thus, by integrating in the six coordinates $x^1, x^2,...,x^6$, we find the four-dimensional Lagrangian on the junction 
\begin{align}
{\cal L}^F_{4d} & =\overline{\tau^{junc}_{0...0}}\Gamma^{\mu}\partial_{\mu}\tau^{junc}_{0...0}+\sum_{n_1...n_6}\overline{\tau^{junc}_{n_1...n_6}}(\Gamma^{\mu}\partial_{\mu}-E^{n_1...n_6}_{junc})\tau^{junc}_{n_1...n_6}+ \nonumber \\ 
                & +\sum_{l_1...l_6}\sum_{m_1...m_6}\sum_{n_1...n_6}g\xi^{junc}_{l_1...l_6}\overline{\tau^{junc}_{m_1...m_6}}\tau^{junc}_{n_1...n_6}.
\label{eq3.20}
\end{align}
Since the system allows two bound states, then we must have $2^n$ supersymmetric partners. As shown above, for $k=1$, we recover the results obtained in \cite{BBF2009}, that is, the case where we have $2^6=64$ four-dimensional scalars $\xi^{junc}_{n_1...n_6}(y^{\mu})$ and $2^6=64$ four-dimensional Dirac fermions $\tau^{junc} _{n_1...n_6}(y^{\mu})$ living on the junction. The first term of the Lagrangian describes massless four-dimensional fermions, while the second term describes massive fermions. The Yukawa couplings are controlled by the constant $g$, which is obtained by integrating the Yukawa couplings in the six extra dimensions.

According to the spectrum given by  Eqs.~\eqref{eq3.11}-\eqref{eq3.18}, for the case $k=1$, in \eqref{eq3.20} we have $2^6$ bound states on the junction. The first bound state corresponds to the zero mode, with zero energy, and the others correspond to the massive modes with energy $m=\sqrt{3}$ given by: 6 states with energy $m$, 15 states with energy $\sqrt {2}m$, 20 states with energy $\sqrt{3}m$, 15 states with energy $\sqrt{4}m$, 6 states with energy $\sqrt{5}m$ and a single state with energy $\sqrt{6}m$, which leads us to the following distribution that respects \eqref{distribution}:
\begin{align}
(N_f,m)= & \{(1,0),(6,m),(15,\sqrt{2}m),(20,\sqrt{3}m), \nonumber \\
         & (15,\sqrt{4}m),(6,\sqrt{5}m),(1,\sqrt{6}m)\}.
\label{eq3.20a}
\end{align}
This shows us that the fermions follow some kind of mass hierarchy in the Lagrangian, which is given by
\begin{align}
{\cal L}^F_{4d} & =\bar{\tau}^{(0)}_0\Gamma^{\mu}\partial_{\mu}\tau^{(0)}_0+\sum^6_{s=1}\sum^{N_s}_{n=1}\bar{\tau}^{(s)}_{n}(\Gamma^{\mu}\partial_{\mu}-\sqrt{s}m)\tau^{(s)}_n+ \nonumber \\ 
                & +\sum_{l,l'}\sum_{m,m'}\sum_{n,n'}g_{l'lm'mn'n}\xi^{(l')}_l\bar{\tau}^{(m')}_m\tau^{(n')}_n,
\label{eq3.64}
\end{align}
where $N_1=6$, $N_2=15$, $N_3=20$, $N_4=15$, $N_5=6$, $N_6=1$ and $l'$, $m'$, $n'=0, 1, ..., 6$.


Let us now consider the opposite extreme case, i.e., the limit where $k \rightarrow 0$. At this point, we see that the energy states dependent on $k$ will suffer modifications. So, for our current case, the energy distributions are given in the form: 5 states with energy $m$, 10 states with energy $\sqrt{2}m$, 10 states with energy $\sqrt{3}m$, 5 states with energy $\sqrt{4}m$ and 1 state with energy $\sqrt{5}m$. Thus, by including the zero mode, we have now a total of 32 bound states. 

Thus in summary, for $N=6$ intersecting domain walls, we have $2^5 \times 2^k$ bound states in the four-dimensional Lagrangian, 
that means as $k \rightarrow 0$, we have $2^5 \times 2^0=2^5=32$ bound states. 
The energy distribution for the present case leads us to
\begin{align}
(N_f,m)=& \{(1,0),(5,m),(10,\sqrt{2}m),(10,\sqrt{3}m), \nonumber \\
        & (5,\sqrt{4}m),(1,\sqrt{5}m)\}.
\label{eq3.21}
\end{align}
Now the fermion mass hierarchy is written as
\begin{align}
{\cal L}^F_{4d} & =\bar{\tau}^{(0)}_0\Gamma^{\mu}\partial_{\mu}\tau^{(0)}_0+\sum^5_{s=1}\sum^{N_s}_{n=1}\bar{\tau}^{(s)}_{n}(\Gamma^{\mu}\partial_{\mu}-\sqrt{s}m)\tau^{(s)}_n+ \nonumber \\ 
                & +\sum_{l,l'}\sum_{m,m'}\sum_{n,n'}g_{l'lm'mn'n}\xi^{(l')}_l\bar{\tau}^{(m')}_m\tau^{(n')}_n,
\label{eq3.22}
\end{align}
where $N_1=5$, $N_2=10$, $N_3=10$, $N_4=5$, $N_5=1$ and $l'$, $m'$, $n'=1, 2, ..., 5$. The maximal entropy of the $d=N+3$-dimensional system is given by \eqref{states-number-3}, i.e.,
\begin{equation}
S\propto\frac{L^{N+2}}{l_P^{N+2}}, 
\label{states-number-3-0}
\end{equation}
so it is clear why the number of states increases with the number of intersecting branes.

\begin{figure}[h!]
\begin{center}
  \includegraphics[scale=0.8]{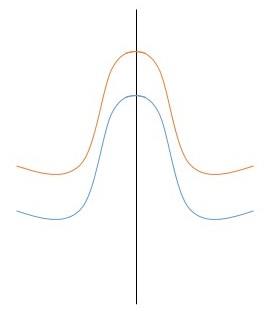}
  \caption{In the case where $k=1$, the fermions are localized.}
  \label{fig.5}
  \end{center}
\end{figure}

\begin{figure}[h!]
\begin{center}
  \includegraphics[scale=0.7]{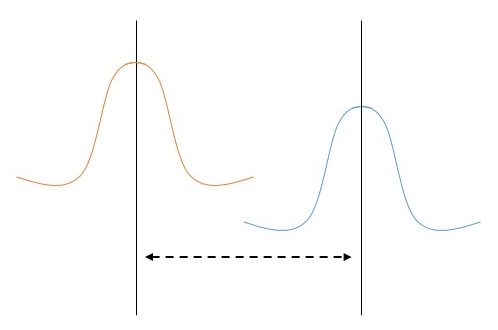}
  \caption{For $k\rightarrow 0$, the fermions are delocalized. The dashed arrow indicates fermions separation along the width of the domain walls.}
  \label{fig.6}
  \end{center}
\end{figure}

We can also understand the ``loss" of states as  $k\to0$ by considering that the fermions are now delocalized (field smeariness.) That is, for $k\rightarrow0$, there is a certain number of fermions that cannot be localized on the junction at the same positions along its thickness (see Fig.~\ref{fig.5} and \ref{fig.6}.) Alternatively, the smeariness (`decompactification') along the coordinate $x^1$ implies that we can rewrite the four-dimensional Lagrangian (\ref{eq3.20}) as 
\begin{align}
{\cal L}^F_{4d} &\! =\!\int dx^1\Big[\: \overline{{\mathcal T}^{junc}_{0...0}(y^\mu,x^1)}\Gamma^{m}\partial_{m}{\mathcal T}^{junc}_{0...0}(y^\mu,x^1)+\!\!\!\!\sum_{n_2...n_6}\overline{{\mathcal T}^{junc}_{n_2...n_6}(y^\mu,x^1)}\Big(\Gamma^{m}\partial_{m}-E^{n_2...n_6}_{junc}+2\lambda_1\phi^1(x^1)\Big){\mathcal T}^{junc}_{n_2...n_6}(y^\mu,x^1)\nonumber \\ 
                & +\sum_{l_2...l_6}\sum_{m_2...m_6}\sum_{n_2...n_6}\Xi^{junc}_{l_2...l_6}(y^\mu,x^1)\overline{{\mathcal T}^{junc}_{m_2...m_6}(y^\mu,x^1)}{\mathcal T}^{junc}_{n_2...n_6}(y^\mu,x^1)\Big].
\end{align}
Notice that now the indices $n_1, l_1, m_1$ do not enter in the counting of bound states. This also explains the lowering in the number of bound states as $k\to0$, since on the opposite side, i.e., at $k\to1$, compactification leads to more bound states. The massless fermions can also have their positions given by the zeroes of $M_{\{n_i\}}=2\lambda_1\phi^1(x^1)$, where $M_{\{n_i\}}\equiv E^{n_2...n_6}_{junc}$. Thus, we can easily establish a connection among the masses and 
particular positions $x^1_{\{n_i\}}=r_{\{n_i\}}\sim M_{\{n_i\}}\Delta_1^2$.  Here, we have used the typical domain wall profiles (\ref{eq3.4}), by considering  $\phi^1(x^1)\sim a_1\,x^1/\Delta_1$, since this domain wall is thicker than the other ones with profile sufficiently thin, given by $\phi^i(x^i)\sim a_i\,{\rm signum}(x^i)$, for $i\geq2$. We might also highlight this relationship as follows
\begin{equation}
M_{\{n_i\}}\sim \frac{x^1_{\{n_i\}}}{\Delta_1^2}.
\end{equation}
Interestingly, this in accord with the mass of strings attached to different branes whose masses are in direct relation with the inter-brane distance $\delta x^1$ and string tension  $ l_s^{-2}$, here analogous to $\Delta_1^{-2}$ --- see also Eq.~(\ref{Mstring}). This also reveals that the domain wall thickness can be considered as the description of a stack of $K$ branes in the continuum limit, i.e., $K\to\infty$ and $\delta x^1 \to 0$ fixed at a compactification radius, that we assume to be $\Delta_1=K\delta x^1$. This scenario has been recently addressed in the context of fractons from branes \cite{Geng:2021cmq}.

\section{Yukawa coupling and the proton decay}\label{sec03}

 In this section, we shall  restrict our number of intersections and dimensions. We shall assume $N=3$ intersecting 5-branes that join together in a seven-dimensional spacetime to form a 3-brane as a junction whose four-dimensional world-volume plays the hole of the spacetime of our Universe.

Let us now compute the Yukawa coupling and the proton decay \cite{AS2000}. 
The Yukawa coupling stands for the interaction between both fermion and scalar fields. Here, the coupling will be due to the orthogonal intersection of three 5-branes in the seven-dimensional spacetime. Let us take this intersection with one of the 5-branes being significantly thicker than the others. Also, for light fields, let us consider the zero mode at different positions along a 5-brane thickness.

Thus, the Yukawa coupling in the four-dimensional junction spacetime is given by
\begin{align}
{\cal L}_{Yukawa} & = 
\xi_{000}\bar{\tau}_{000}\tau_{000}\int{\psi^{000}\times \psi^{000} \times \psi^{000}\,  dx^1dx^2dx^3} \nonumber\\
                  & =g\, \xi_{000}\bar{\tau}_{000}\tau_{000},
\label{eq4.1}
\end{align}
where we consider
\begin{eqnarray}
\!\!\!\!\!\psi^{000}=C_1{\rm sech}^2{\left(\frac{x^1}{\Delta_1}\right)}\times {\rm sech}^2{\left(\frac{x^2}{\Delta_2}\right)}\times {\rm sech}^2{\left(\frac{x^3}{\Delta_3}\right)}, \:\:\:\: \Delta_1\simeq\frac{1}{a_1},\: \Delta_2\simeq\frac{1}{a_2},\: \Delta_3\simeq\frac{1}{a_3}, \:\:\:\: a_1=k\, a_2,\; a_2=a_3=a,
\label{eq4.2}
\end{eqnarray}
so that $\Delta_{1,2,3}$ are interpreted as the width of the three 5-branes.

To understand the hierarchy problem we had considered that the leptons are distributed in different locations along the width of one of the branes. Let us admit that {\it the width of the brane extends in the $x^1$ direction}, so that one lepton is in the $x^1$ position, and the other one is in the $x^1-r$ position. Since each contribution of the integration in the other coordinates related to sufficiently thin domain walls is $\sim 1$, then we simply find at the junction  
\begin{eqnarray}
g&\sim&\int{dx^1\:{\rm sech}^4{\left(\frac{x^1}{\Delta_i}\right)}\times {\rm sech}^2{\left(\frac{x^1-r}{\Delta_i}\right)}},
\label{eq4.3}
\end{eqnarray}
where we have assumed $\Delta_i\sim 1/a, i\geq2$ as the thickness associated with the fundamental scale $a$ at the junction --- see Eq.(\ref{eq4.2}) --- i.e, $\Delta_i$ is the smallest width related to the other $i$-th intersecting branes.
Thus, for $ar\simeq 8$, or better, $r\simeq 8\Delta_2\lesssim \Delta_1$, i.e., $k\lesssim1/8$,
we have $g_e\simeq 1.2\times 10^{-6}$, which is the coupling for the electron --- See \cite{dEnterria:2021xij} for very recent discussions. For $r\gg \Delta_2$ we have indeed an explicit form given by $g\sim e^{-2r/\Delta_2}$. 

The proton decay is obtained through quadrilinear operators between quarks and leptons (interactions involving four fermions.) Thus, we follow this prescription by considering Majorana bilinear $\Psi_1^T C_n \Psi_2$ in $n$ spacetime dimensions. 

In order to get the proton decay we must integrate the operator $(\Psi_1^T C_n \Psi_2)^\dagger(\chi_1^{cT} C_n \chi_2^c)$, where $\chi^c$ is the charge-conjugate of the spinor $\chi$. Using the same definitions of spinors on the previous junction, and applying the product between three spinors (quarks) in the $x^1$ position and one spinor (lepton) in the $x^1-r$ position, we finally get the proton decay in the form
\begin{eqnarray}
\delta&\sim&\int{dx^1\:{\rm sech}^6{\left(\frac{x^1}{\Delta_i}\right)}\times {\rm sech}^2{\left(\frac{x^1-r}{\Delta_i}\right)}}.
\label{eq4.4}
\end{eqnarray}
Following the same previous steps, $\Delta_i\sim 1/a, i\geq2$,  for $ar\simeq 40$, that means $r\simeq 40\Delta_2\lesssim \Delta_1$, i.e., $k\lesssim1/40$, we find $\delta\sim 1.15\times 10^{-34}$, which is an acceptable value in the 1TeV scale \cite{AS2000}.

\begin{figure}[h!]
\begin{center}
  \includegraphics[scale=0.3]{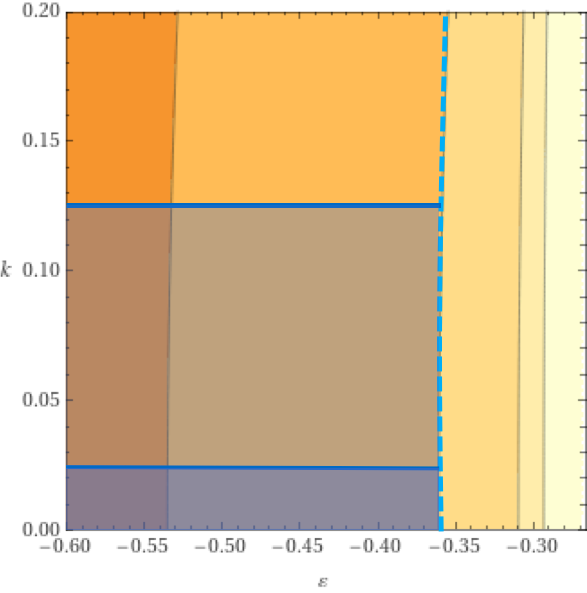}
  \caption{The contour plot of $\Lambda(k,\varepsilon)$ given in Eq.~(\ref{eq2.9}) for $0\leq k\leq 0.2$. In the contour along the dashed-blue curve $\Lambda=0$, whereas in the region inside, $\Lambda<0$ (i.e., the $AdS$ radius $\sim 1/|\Lambda|$ becomes smaller and opens up the regime of higher energies.) The regions below the thicker and thinner horizontal lines corresponds to electron Yukawa coupling $(k\leq 1/8)$) and proton decay $(k\leq 1/40$), respectively. Thus the darker region below the thinner horizontal line restricts the parameters to attend both Yukawa coupling and proton decay since it corresponds to the maximum brane thickness $\Delta_1\gtrsim 40\Delta_i$, where $\Delta_i$ is the smallest width related to the other $i$-th intersecting branes, with $i\geq2$. Along the vertical contours $\Lambda$ does not change.}
  \label{fig.1.3.2}
  \end{center}
\end{figure}

In Fig.~\ref{fig.1.3.2} the darker region below the thinner horizontal line restricts the parameters to attend both Yukawa coupling and proton decay since it corresponds to the maximum brane thickness $\Delta_1\simeq 40\Delta_i$. Thus, because the extra dimension is large enough in this regime, this is the region where there is no need of symmetry to explain  hierarchies \cite{AS2000}. On the other hand, above the thicker horizontal line, symmetries are needed, as long as the vertical dashed curve is never crossed, i.e., $\Lambda>0$ is never achieved. For $r\simeq \ell \Delta_i$, and $\ell=1,2,..,8,...,40$, the heavier leptons stand for smaller values of $\ell$.

Now because of the parameters $k$ and $\varepsilon$ playing the role together in the theory, the Figs.~(\ref{fig.1.3.1}) and (\ref{fig.1.3.2}) show a contour where there is a richer possibility of o maintaining the junction stability ($\varepsilon<0$) along a curve where the bulk cosmological constant $\Lambda$ is zero. Thus, we should address this issues more suitably as follows. 
Since the cosmological constant plays the role of the radius of compactifications $L\sim1/\Lambda$ due to gravity that warps the spacetime, then $L\to\infty$ along the aforementioned contour, which means that the junction is now embedded in a Minkowski spacetime. As a consequence the gravitational potential behavior on the junction is higher dimensional rather than four-dimensional. This is because at the junction the Newtonian potential for $r\gg L$ is given by \cite{ADDK2000}
\begin{eqnarray}
V(r)\sim \frac{G M}{r}, \qquad G\sim\frac{G_{(4+N)}}{L^N}
\end{eqnarray}
whereas for $r\ll L$ is
\begin{eqnarray}
V(r)\sim \frac{G_{(4+N)}M}{r^{N+1}}.
\end{eqnarray}
In our case, we are assuming $N=3$ intersecting  5-branes, thus for $r\ll L$ we find the seven-dimensional Newtonian potential
\begin{eqnarray}
V(r)\sim \frac{G_{(7)}M}{r^{4}}.
\label{7d-N}
\end{eqnarray}
However as $L\to\infty$, as it happens along the dividing line, i.e., the dashed-blue curve in Fig.~\ref{fig.1.3.2}, the Newtonian formula (\ref{7d-N}) is true in any scale.


\section{Conclusions}\label{sec04}

We have considered the braneworlds scenario where we dealt with intersecting branes with different thickness. For six intersecting 8-branes at ten-dimensional spacetime, the resulting junction is four-dimensional and describes localized fermion fields with a zero mode  and a specific distribution of massive modes. As one of these branes is sufficiently thicker than the others, only half of massive states is preserved. This is then interpreted as an effect of the delocalization of the fermion field along the brane thickness, i.e., now fermion fields can be localized at different positions in the extra dimension according to their masses. More specifically, we applied this effect to compute the Yukawa coupling and the proton decay for the intersection of three 5-branes in seven-dimensional spacetime. By combining the contour map of the bulk cosmological constant as a function of two parameters, we can show that the proton decay can impose important bounds on the parameters.

{\acknowledgments} 

We would like to thank CNPq, CAPES and CNPq/PRONEX/FAPESQ-PB (Grant no. 165/2018),
for partial financial support. FAB acknowledges support from CNPq (Grant no. 312104/2018-9).

\end{document}